\documentclass{aa}
\usepackage{graphicx}

\newcommand{\braref}	[1]	{(\ref{#1})}
\newcommand{\bracite}	[1]	{(\cite{#1})}
\newcommand{\nuc}	[2]	{\hbox{$^{#1}$}\hbox{\rm{#2}}}
\newcommand{\beq}		{\begin{equation}}
\newcommand{\eeq}		{\end{equation}}
\newcommand{\beqarr}		{\begin{eqnarray}}
\newcommand{\eeqarr}		{\end{eqnarray}}
\newcommand{\beqarrs}		{\[\begin{array}{l}}
\newcommand{\eeqarrs}		{\end{array}\]}
\def\const			{{\rm const}}
\def\etal			{{et~al.}}
\def\PROMETHEUS			{{\sc prometheus}}
\def\vzero{
{\mathchoice {\hbox{$\sf\textstyle 0\kern-0.4em 0$}}
{\hbox{$\sf\textstyle 0\kern-0.4em 0$}}
{\hbox{$\sf\scriptstyle 0\kern-0.3em 0$}}
{\hbox{$\sf\scriptscriptstyle 0\kern-0.2em 0$}}}}
\def\AA				{A\&A}
\def\JCP			{J. Comput. Phys.}
\begin{document}
%%%%%%%%%%%%%%%%

\thesaurus{03(02.08.1; 02.14.1; 03.13.4; 08.19.4)}

\title{The Consistent Multi-fluid Advection method}

\subtitle{}

\author{
        T. Plewa
        \thanks{{\it Present address:\/}
        Nicolaus Copernicus Astronomical Center, Bartycka 18,
        00716 Warsaw, Poland; e-mail: tomek@camk.edu.pl}
   \and E. M\"uller
       }

\offprints{T. Plewa}

\institute{Max-Planck-Institut f\"ur Astrophysik,
           Karl-Schwarzschild-Stra{\ss}e 1,
           Postfach 1523,
           85740 Garching b.\ M\"unchen, Germany \\
           e-mail: (tomek,ewald)@mpa-garching.mpg.de
          }

\date{Received \quad\quad\quad ; accepted \quad\quad\quad }

\maketitle
\markboth
{T. Plewa \& E. M\"uller: Consistent Multi-fluid Advection}
{T. Plewa \& E. M\"uller: Consistent Multi-fluid Advection}

%%%%%%%%%%%%%%%%
\begin{abstract}
%%%%%%%%%%%%%%%%

Simple modifications for higher-order Godunov-type difference schemes
are presented which allow for accurate advection of multi-fluid flows
in hydrodynamic simulations. The constraint that the sum of all mass
fractions has to be equal to one in every computational zone
throughout the simulation is fulfilled by renormalizing the mass
fractions during the advection step. The proposed modification is
appropriate for any difference scheme written in conservation
form. Unlike other commonly used methods it does not violate the
conservative character of the advection method.  A new steepening
mechanism, which is based on modification of interpolation profiles,
is used to reduce numerical diffusion across composition
discontinuities. Additional procedures are described, which are
necessary to enforce monotonicity. Several numerical experiments are
presented which demonstrate the capability of our Consistent
Multi-fluid Advection (CMA) method in case of smooth and
discontinuous distributions of fluid phases and under different
hydrodynamic conditions.  It is shown that due to the reduced
diffusivity of the proposed scheme the abundance of some heavy
elements obtained from hydrodynamic simulations of type II supernova
explosions can change by a factor of a few in the most extreme cases.

%%%%%%%%%%%%%%
\end{abstract}
%%%%%%%%%%%%%%

%%%%%%%%%%%%%%%%%%%%%%
\section{Introduction}				\label{s:intro}
%%%%%%%%%%%%%%%%%%%%%%

One of the most important factors determining the quality of a
numerical algorithm is its robustness. In the simplest case it can be
regarded as a property of the scheme to provide the result at minimum
cost once the accuracy is specified. In numerical hydrodynamics much
effort has been spent during the last two decades on improving
advection schemes in such a way that they do not only provide high
accuracy in regions of smooth flows but also resolve flow
discontinuities (shocks and contact discontinuities) with a minimum
number of discrete grid zones. In the past a variety of numerical
experiments were conducted to compare the overall quality of solutions
obtained with the help of different advection schemes which provided
an understanding of the advantages and disadvantages of already
available or newly proposed algorithms (Colella \& Woodward
\cite{cw84}, Carpenter \etal\ \cite{c+90}, Fryxell \etal\
\cite{fma91}, Yang \& Przekwas \cite{yp92}, Stone \& Norman
\cite{sn92}, Steinmetz \& M\"uller \cite{sm93}, Kang \etal\
\cite{k+94}).

For our numerical experiments we have used the Piecewise Parabolic
Method (PPM) of Colella \& Woodward \cite{cw84}; hereafter CW) to
study the evolution of multi-fluid flows with strong discontinuities
and stiff source terms. In theoretical astrophysics the PPM method has
been used to study a range of hydrodynamic phenomena like stellar
collisions (Ruffert \& M\"uller \cite{rm90}, Frolov \etal\
\cite{f+94}), evolution of supersonic jets (Balsara \& Norman
\cite{bn92}, Basset \& Woodward \cite{bw95}), large-scale structure
formation in cosmology (Bryan \etal\ \cite{b+94}), interaction of
stellar winds in massive close binaries (Stevens \etal\ \cite{sbp92}),
and the stability of radiative shock waves (Strickland \& Blondin
\cite{sb95}). Numerical experiments of CW and Yang \& Przekwas
\bracite{yp92} clearly demonstrated the superiority of the PPM scheme
among several modern advection schemes.

A particularly interesting and challenging astrophysical problem
involving multi-fluid flow (and one of our numerical experiments; see
section~3.4) is the simulation of mixing in supernova envelopes.
Mixing occurs because the non-steady propagation of the shock wave
formed after core collapse gives rise to Rayleigh-Taylor instabilities
(for a recent review, see e.g., M\"uller \cite{m98}).  The first (2D)
simulations of mixing involving ten separate fluids were performed by
Arnett \etal\ \bracite{afm89}.  They computed the propagation of the
supernova shock wave, which was artificially created by a ``point''
explosion, through the envelope of a realistic stellar model.  Better
resolved and more detailed simulations were later performed by Fryxell
\etal\ \bracite{fma91}\ and M\"uller \etal\ \bracite{mfa91}.  They
identified the Rayleigh-Taylor unstable regions as being associated
with discontinuities in the chemical composition in the envelope of
the progenitor star. The simulations were performed with the PPM-based
hydrodynamic code {\PROMETHEUS}, which keeps track of different nuclear
species by solving a set of additional continuity equations (see
Fryxell \etal\ \cite{fma89}).

In the case of single-fluid advection the problem of diffusion across
contact discontinuities plays a crucial role and provides a simple
test case for studying mixing between different fluids in numerical
simulations. In multi-fluid flows both chemical and contact
discontinuities may be present. Although, as we shall see later, there
exist important differences between both kinds of discontinuities, we
nevertheless can profit from our experience of modeling contact
discontinuities when dealing with composition discontinuities. In this
context we point out that Fryxell \etal\ (\cite{fma89}) demonstrated
that the advection of a contact discontinuity is simulated better with
a PPM scheme than with any other scheme they considered in their
study.

Mixing of different fluids cannot be ignored if the chemical
composition plays an important role in the hydrodynamic evolution. For
example, in case of a realistic equation of state the total gas
pressure is calculated as the sum of partial pressures exerted by each
kind of species. More complex physical processes, emission from an
optically thin medium (radiative cooling) or absorption of radiation,
are strongly sensitive to changes in chemical composition, especially
to changes in the heavy element abundances. Last but not least, the
process of nuclear burning, to which we will pay special attention
later in this paper, directly depends on the amount and type of
nuclear fuel. In this particular case, mixing of different nuclear
species due to numerical diffusion can substantially affect the final
chemical composition or even the overall dynamics of the flow (for a
recent review, see e.g., M\"uller \cite{m98}).

The paper is organized as follows. In Sect.\ \ref{s:method} we briefly
describe the basic components of the PPM scheme and some specific
features implemented into the \PROMETHEUS\ version used in our
numerical experiments. We then give a detailed description of the new
consistent multi-fluid advection method. In section \ref{s:results} we
present the results of our test simulations. A discussion of the
results is contained in Sect.\ \ref{s:conclusions}.

%%%%%%%%%%%%%%%%%%%%%%%%%%
\section{Numerical method}			\label{s:method}
%%%%%%%%%%%%%%%%%%%%%%%%%%

In what follows, we consider only the {\em Direct Eulerian}
formulation of the PPM scheme as implemented in the \PROMETHEUS\ code
(Fryxell \etal\ \cite{fma89}). Most of the presented results are valid
for and can efficiently be implemented in codes based on the {\em
Lagrangian with remap} approach, too.  Both versions of the PPM method
belong to the class of shock-capturing methods which are characterized
by high-order spatial and temporal accuracy. In such schemes flow
discontinuities are treated as solutions to the hydrodynamic equations
(Riemann problem) and are represented by sharp (1 to 2 zones wide) and
oscillation free profiles of hydrodynamic variables. There is no need
for adding large amounts of artificial viscosity for shocks to obtain
correct post-shock states. \PROMETHEUS\ uses a Strang-type directional
splitting (Strang \cite{s68}) for simulation of multidimensional
flows.

For given initial data and boundary conditions each hydrodynamic step
of the PPM scheme begins with construction of the interpolants
approximating the distribution of flow variables inside each grid
zone. The initial parabolic profiles are subsequently modified
according to local flow conditions: density profiles are made steeper
near contact discontinuities and the distributions of all variables
are somewhat flattened near shocks in order to reduce high-frequency
post-shock oscillations.  Afterwards, monotonicity constraints are
imposed on the interpolation profiles to avoid unphysical solutions.
The monotonized profiles are used to calculate initial data for the
Riemann problem at each zone interface by averaging the monotonized
parabolae over the domain of dependence of the zone interface.  The
left and right states at the interface obtained in this way define the
input data for the nonlinear Riemann problem, which is solved
iteratively. The solution of the Riemann problem provides average
hydrodynamic state variables at the zone interface, which are used to
compute the fluxes for the advection step, whereby the hydrodynamic
variables are advanced to the new time level.

For simulations of mixing in supernova explosions the basic PPM
algorithm was modified to allow for advection of several nuclear
species.  The hydrodynamic state vector was extended by adding mass
fractions for each of the species as were the left and right states
used as input for the Riemann problem. The effective states at zone
interfaces are obtained by averaging the mass fraction profiles over a
properly chosen domain of dependence for the zone interface. The
interpolation step for the mass fractions also includes steepening and
flattening procedures followed by a monotonization step. The mass
fractions do not enter the Riemann problem. They are treated as
passive scalars and are advected with the flow depending on the upwind
state as determined by the average velocity obtained from the solution
of the Riemann problem.

\PROMETHEUS\ includes the handling of a general equation of state
using the approach of Colella \& Glaz \bracite{cg85}.  Gravitational
forces are included in the calculation of the effective states
entering the Riemann problem and by a separate acceleration step at
the end of each hydrodynamic sweep. Operator splitting is also used to
couple nuclear burning and hydrodynamics. The nuclear reaction network
is solved using a multidimensional Newton iteration (M\"uller
\cite{m86}). A more detailed description of the implementation of
\PROMETHEUS\ can be found in Fryxell \etal\ \bracite{fma89} and
M\"uller \bracite{m98}.

%%%%%%%%%%%%%%%%%%%%%%%%%%%%%%%%%%
\subsection{Multi-fluid advection}
%%%%%%%%%%%%%%%%%%%%%%%%%%%%%%%%%%

We consider the one-dimensional initial-boundary value problem
\[
\partial_{t}\vec{U} + \partial_{x}\vec{F}(\vec{U}) = \vec{G},
\]
with boundary data $\vec{U}(x=x_{\rm L},t)$ and $\vec{U}(x=x_{\rm
R},t)$, where \vec{F}, \vec{U} and \vec{G} are the flux vector, the
state vector and the vector of source terms, respectively. In case of
the Euler equations for multi-fluid ideal hydrodynamics the state
vector is
\[
\vec{U}=\vec{U}(x,t)=\pmatrix{\rho \cr \rho u \cr \rho E \cr \rho X_{n}}
\, , \quad
\vec{F}(\vec{U}) =\pmatrix{\rho u \cr \rho u^2+p \cr (\rho E + p) u \cr 
                           \rho X_{n} u},
\]
where $\rho, u, E = e + u^2/2, e, p, X_n$ and ${\rho}X_n$ are the
total gas density, the velocity, the specific total energy, the
specific internal energy, the pressure and the mass fraction and the
partial density of the $n$-th fluid, respectively.  The closure
relation for the above system is provided by the equation of state,
$p=p(\rho, e, \vec{X})$, where \vec{X} is the vector of mass
fractions.

Let the number of different fluid phases be $N_X$ and the mass
fraction of the $n$-th fluid inside zone $i$ be $X_{i,n}$.  Then the
following relation must hold for $t{\ge}t_{0}$,
\beq
\sum\limits_{n=1}^{N_X}{X_{i,n}} = 1.
\label{e:sumx}
\eeq
%

%%%%%%%%%%%%%%%%%%%%%%%
\subsection{FMA method}
%%%%%%%%%%%%%%%%%%%%%%%

As it has been observed first by Fryxell \etal\ \bracite{fma89} using
\PROMETHEUS\ and by Larrouturou \bracite{l91}, the nonlinear
character of higher-order Godunov-type schemes is the primary reason
for the violation of \braref{e:sumx}. During a simulation it is not
guaranteed that in such schemes the sum of the mass fractions inside
each zone remains equal to unity even if both the underlying advection
scheme is conservative and the total mass of each fluid (summed over
the whole grid) remains constant to within machine accuracy (see also
M\"uller \cite{m98}).

This failure of high-order schemes can be understood when one realizes
that interpolation profiles are constructed independently for each
mass fraction. Thus their sum can take an arbitrary value.  We notice
that this problem has a predominantly local character: (i) it is most
important in regions where changes in composition are substantial, and
(ii) condition \braref{e:sumx} can be violated for any subvolume of a
zone (e.g. for the zone of dependence).

In practice, condition \braref{e:sumx} is usually enforced by applying
a simple renormalization of the mass fractions after each step. This
procedure, however, not only lacks any formal justification but it
also leads to large systematic errors in the mass fractions of the
least abundant species.  It also violates the conservative character
of the scheme.

One possible solution to this problem is to modify the interpolation
step in such a way that deviations of the sum of the mass fractions
from unity will remain small inside each zone. According to the
procedure proposed by Fryxell \etal\ \bracite{fma89}, which we will
refer to as FMA, this can be obtained by calculating the sum of the
mass fractions at the zone interface and to flatten the interpolation
profiles for that zone totally, if
\beq
{\Delta}_{\Sigma}^{\#} = \left| \sum\limits_{i=1}^{N_X} 
                                \vec{X}_{i}^{\#} - 1    \right| 
                       \ > \ \varepsilon_\Sigma \, ,
\label{e:delsig}
\eeq
with some predefined threshold value $\varepsilon_\Sigma$. Here, and
in what follows, we take expressions involving $\#$ to mean $\#$ equal
to either $+$ or $-$, where $+$ and $-$ refer to the right and left
interface of the $i$-th zone, respectively.

In their original calculations Fryxell \etal\ \bracite{fma89} used
\PROMETHEUS\ supplemented with FMA (with $\varepsilon_\Sigma =
10^{-7}$, while a less restrictive value of $\varepsilon_\Sigma =
10^{-3}$ seems to be acceptable for most applications). Although FMA
acts only locally, in the long term it affects large regions of the
grid due to its high diffusivity which effectively reduces the spatial
accuracy of advection of species to first order. Consequently, one has
to expect a large amount of mixing of fluids especially near
composition interfaces.

%%%%%%%%%%%%%%%%%%%%%%%%
\subsection{CMA method}				\label{s:cma}
%%%%%%%%%%%%%%%%%%%%%%%%

The Consistent Multi-fluid Advection (CMA) method retains the high
accuracy of the PPM advection scheme for mass fractions with
constraint \braref{e:sumx} being accurately fulfilled, while any
excessive flattening of interpolation profiles (as seen in case of
FMA) is avoided.

\noindent
The advection step can be written as,
\beqarr
\lefteqn{
\sum\limits_{i=1}^{N} {\rho}_{i}\vec{X}_{i}(t+\Delta t){\Delta V}_{i}
= } && \nonumber \\
&&\null
\sum\limits_{i=1}^{N} {\rho}_{i}\vec{X}_{i}(t         ){\Delta V}_{i}
- {\Delta t}\sum\limits_{i=1}^{N}
( A_{i}^{+}\vec{F}_{i}^{+} - A_{i}^{-}\vec{F}_{i}^{-} )
\label{e:advx}
\eeqarr
where $\vec{F}_{i}^{\#}$ (with $\# \in \{+,-\}$) is the numerical flux
vector across the zone interface of zone $i$.  $A_{i}^{\#}$ is the
area of the zone interface, ${\Delta V}_{i}$ is the zone volume and
$\Delta t$ the size of the time step.

Since the PPM scheme is conservative,
\beq
\vec{F}_{i}^{-} = \vec{F}_{i-1}^{+},
\label{e:flxlr}
\eeq
the following condition holds for each component of $\rho\vec{X}$
separately:
\beq
\sum\limits_{i=1}^{N} \rho_{i} X_{i,n}(t) {\Delta V}_{i} = \const.
\label{e:fsumconst}
\eeq
We seek for a set of numerical fluxes, $\vec{{\mathcal{F}}}_{i}^{\#}$,
which satisfy conditions \braref{e:sumx} and \braref{e:fsumconst}
simultaneously. In general, the total mass flux at a zone interface,
${\rho}_{i}^{\#} u_{i}^{\#}$, computed from the sum of the mass fluxes
of individual species is not necessarily equal to the total mass flux
computed with the (total) mass density $\rho_{i}^{\#}$:
\beq
\sum_{n=1}^{N_X} {F}^{\#}_{i,n} 
=
\sum_{n=1}^{N_X} {\rho}_{i}^{\#} u_{i}^{\#} {X}^{\#}_{i,n} \neq
                 {\rho}_{i}^{\#} u_{i}^{\#}.
\label{e:mfx}
\eeq
This inconsistency is the reason why any higher-order Godunov-type
advection scheme in which the interpolation step for the mass
fractions (or partial densities) is not appropriately constrained will
violate condition \braref{e:sumx}. The inconsistency can be avoided
when scaling the original partial mass fluxes ${F}_{i,n}^{\#}$ in
such a way that their sum is exactly equal to the total mass flux.

We request the modified partial mass fluxes of zone $i$,
${\mathcal{F}}_{i,n}^{\#}$ (which depend on the modified mass
fractions at the interface ${\mathcal{X}}^{\#}_{i,n}$) to be consistent
with the advection of the total mass, i.e.
\beq
\sum_{n=1}^{N_X} {\mathcal{F}}^{\#}_{i,n} 
\equiv
\sum_{n=1}^{N_X} {\rho}_{i}^{\#} u_{i}^{\#} {\mathcal{X}}^{\#}_{i,n} =
                 {\rho}_{i}^{\#} u_{i}^{\#}.
\label{e:nfx}
\eeq
Hence, we modify the original partial mass flux vector using the
simple scaling operation
\beq
\vec{{\mathcal{F}}}_{i}^{\#} = {\varphi}_{i}^{\#}\vec{F}_{i}^{\#},
\label{e:fscale}
\eeq
in such a way that it becomes consistent with the continuity equation:
\beq
{\varphi}_{i}^{\#}
\sum_{n=1}^{N_X} {\rho}_{i}^{\#} u_{i}^{\#} {X}^{\#}_{i,n} 
=
                 {\rho}_{i}^{\#} u_{i}^{\#}.
\label{e:ofx}
\eeq
Comparing \braref{e:nfx} with \braref{e:ofx} one obtains
\beq
{\varphi}_{i}^{\#}
\sum_{n=1}^{N_X} {X}^{\#}_{i,n}
=
\sum_{n=1}^{N_X} {\mathcal{X}}^{\#}_{i,n},
\label{e:phidef}
\eeq
i.e. the normalization constant, ${\varphi}_{i}^{\#}$, can be written
in terms of the (unknown) modified mass fractions,
${\mathcal{X}}^{\#}_{i,n}$, as
\beq
{\varphi}_{i}^{\#} =
\frac{\sum_{n=1}^{N_X} {\mathcal{X}}^{\#}_{i,n}}
     {\sum_{n=1}^{N_X}          {X }^{\#}_{i,n}}.
\label{e:fullphi}
\eeq
Since the modified mass fractions are consistent with advection of the
total mass by definition (\ref{e:nfx}), their sum is equal to unity,
i.e.
\beq
\varphi^{\#}_{i} = \frac{1}{\sum_{n=1}^{N_X} X^{\#}_{i,n}}.
\label{e:phi}
\eeq

In practice it is not necessary to explicitly compute the modified
partial mass fluxes (\ref{e:fscale}), but instead one simply
has to scale the original mass fractions ${X}^{\#}_{i,n}$ according to
(\ref{e:phidef}), i.e.
\beq
{\mathcal{X}}^{\#}_{i,n} = {\varphi}_{i}^{\#} {X}^{\#}_{i,n}.
\label{e:xscale}
\eeq
The CMA method does not require any modification of the interpolation
step of the advection scheme.  Furthermore, flux scaling does not
destroy the conservative character of the scheme
(\ref{e:flxlr}). Hence, scaling the average mass fractions
(\ref{e:xscale}) obtained from the solution of the local Riemann
problem by $\varphi^{\#}_{i}$ defined in Eq.\ \braref{e:phi} is
sufficient to satisfy both \braref{e:sumx} and \braref{e:fsumconst}.

In passing we note that instead of scaling the partial mass fluxes one
could, in principle, appropriately adjust the total mass flux, too
(which is equivalent to computing the total mass flux as the sum of
the partial mass fluxes). Although numerical experiments (not
presented here) demonstrate that this method produces results of
comparable quality to the CMA method, we prefer to use the latter
because it preserves the original role of the total gas density in the
PPM method.  Finally, one could discard the continuity equation for
the total density and use instead the partial densities as the primary
variables to calculate the total density whenever needed. We found
this method to give results of overall very poor quality, and being
much more diffusive near contact and composition discontinuities.

Modification \braref{e:xscale} implemented in the original PPM
advection allows us to abandon the FMA scheme, and to retain the high
accuracy of the PPM method. We will refer to this algorithm as
sCMA. There are, however, still some problems left to be solved. In
what follows, we focus on the problem of numerical diffusion across
composition interfaces, and in these cases when the nonlinearity of
the advection scheme becomes too strong to preserve the monotonicity
of the scheme. This will complete the description of all of the
elements constituting the CMA method.

FMA uses a contact steepening mechanism which in PPM is used to limit
diffusion across contact discontinuities only. Removing FMA and using
the full capabilities of the PPM interpolation algorithm for mass
fractions, however, revealed a severe problem. Overshooting occurred
near composition interfaces for the most abundant species. This, in
turn, caused the least abundant species to be evacuated out of the
critical region. Clearly, in its original version the PPM contact
discontinuity detection algorithm cannot be directly used to steepen
distributions of mass fractions. Some additional criteria have to be
used to identify composition interfaces which need to be steepened and
some mechanism has to be devised which limits overshooting.  We point
out that in most cases the FMA method effectively prevents any
steepening of mass fractions profiles, since once any of the fluid
distributions has been steepened, it likely activated the FMA
flattening procedure.

Since we do not expect the geometrical properties of composition
interfaces to be substantially different from that of contact
discontinuities, most of the original steepening algorithm can be
safely used without modification for detecting large gradients in the
fluid's composition. It is very unlikely to find some additional
constraint similar to that originally proposed for detection of
contact discontinuities (CW, Eq.\ 3.2) in order to make the scheme
more selective. Therefore, we will only consider local properties of
composition profiles. One possibility is to associate composition
interfaces with rapid changes in fluid composition. For this purpose
we define a steepness measure for mass fraction profiles
\[
\alpha_{i,n} = \frac{X_{i+1,n}-X_{i-1,n}}{X_{i+2,n}-X_{i-2,n}}.
\]
Additional steepening is applied if $\alpha_{i,n}$ is larger than a
fixed value $\alpha_X$. Since $\alpha_{i,n}$ has a similar meaning as
the parameter $\omega$ used in PPM for calculation of the flattening
coefficients (CW, Eq.\ A8), we set $\alpha_X = 0.75$. To prevent the
steepening of relatively small composition jumps the following
criterion has to be satisfied, too:
\[
|X_{i+1,n}-X_{i-1,n}|\ -\ {\varepsilon_x}\ 
                          {\min}(X_{i+1,n}-X_{i-1,n}) > 0.
\]
We use ${\varepsilon_x} = 0.01$. Furthermore, we do not steepen
composition profiles near extrema. If the zone is located next to an
extremum for which the following criterion has to be fulfilled
\[
(X_{i+2,n}-X_{i+1,n}) \, (X_{i-1,n}-X_{i-2,n}) \le 0,
\]
the steepening coefficient is set to zero. Finally, we do not steepen
abundance profiles inside contact discontinuities, because this gives
rise to enhanced overshooting of the partial densities, $\rho_{i}
X_{i,n}$. After steepening we flatten and monotonize the mass fraction
profiles in the same way as in PPM.

As a remedy for the overshooting near composition discontinuities, we
introduce two additional modifications. Local extrema of the mass
fraction distributions are identified by the criterion
\[
({X_{i+1,n}}-{X_{i,n}}) \, ({X_{i,n}}-{X_{i+1,n}}) < 0.
\]
If the criterion is fulfilled in a zone $i$, the interpolation
profiles are flattened by an additional (constant) amount in the two
zones next to it ($i\pm 1$):
\beq
X_{i\pm 1,n}^{\#} = {w}{X_{i\pm 1,n}} + (1-w){X_{i\pm 1,n}^{\#}}.
\label{e:flatx}
\eeq
We use $w=0.5$.

The second modification is somewhat more complex. For each zone we
calculate two sums 
\beq S_{i}^{\#,\sigma} = \sum_{n=1}^{N_x} \max 
                            \left( 0,\, 
                            \sigma ( {X_{i,n}^{\#}} - {X_{i,n}} ) 
                            \right), 
\quad \sigma \in \{+,-\},
\label{e:dev}
\eeq
at both zone interfaces ($\# \in \{+,-\}$), which give the total
positive $S_{i}^{\#,+}$ and negative $S_{i}^{\#,-}$ deviation between
the values of the mass fractions at the respective zone interface
$X_{i,n}^{\#}$ and the values of the zone averaged mass fractions
$X_{i,n}$.

We then correct the interface values of the mass fractions according
to \braref{e:flatx}. However, instead of the constant flattening
coefficient $w$, we use a variable coefficient $w^{\#}_{i,n}$, which
depends on the positive ($S_{i}^{\#,+}$) and negative ($S_{i}^{\#,-}$)
deviations defined in \braref{e:dev}:
\beq
w^{\#}_{i,n} = s^{\#}_{i,n} \max \left( 0 , \min \left( 1, \, \beta\
\frac{ \Delta^{\#}_{i,max} - \Delta^{\#}_{i,min} }{ \Delta^{\#}_{i,min} }
\right) \right),
\label{e:xmslopes}
\eeq
where
\[
\Delta^{\#}_{i,min} = \min( S^{\#,+}_{i} , S^{\#,-}_{i} ), 
\]
\[
\Delta^{\#}_{i,max} = \max( S^{\#,+}_{i} , S^{\#,-}_{i} ),
\]
and
\[
s^{\#}_{i,n} = 0.5 \, \left| 
              {\rm sign} \left( X^{+}_{i,n} - X^{-}_{i,n}   \right) \ \# \
              {\rm sign} \left( S^{\#,+}_{i} - S^{\#,-}_{i} \right) \right|,
\]
with $\beta=0.25$.

The reasoning behind this procedure is based on the observation that
any variation in the distribution of one fluid component should be
compensated by an appropriate variation of the distribution of at
least one other component. Here, we define two separate groups of
fluid components which show deviations (from the zone average) of the
same sign at the zone interface. Once the sums of the negative
($S^{-}$) and the positive ($S^{+}$) deviations are obtained, we try
to limit their relative difference.  Note that the FMA flattening
criterion is based on the absolute value of the deviation of the sum
$S^{\#,+} + S^{\#,-}$ from unity.

In the CMA method the amount of additional flattening is not constant
but smoothly increases with the relative difference between the two
deviations. Maximum flattening is applied if the relative difference
between the two deviations exceeds $1/\beta$, while no additional
flattening is introduced for $\beta=0$.  Hence, when $\beta > 0$, the
interpolation profiles for that group of species which shows the
largest absolute deviation are modified using the same flattening
coefficient for all species. This reduces the deviation and brings it
closer to that of the other group of species.

In this way we introduce a finite amount of coupling between the
interpolation profiles of two distinct groups of fluid components
rather than trying to adjust the distributions of fluid components
individually.  Since the additional flattening procedure has a
strictly dissipative character and relies on flattening of
interpolation profiles, it is very unlikely that it causes unphysical
solutions.

Finally, we mention that the just described procedure could also be
used to guarantee consistency between the total mass flux and the sum
of the partial mass fluxes. Hence, it could be applied both in the
interpolation and the advection step.

%%%%%%%%%%%%%%%%%
\section{Results}				\label{s:results}
%%%%%%%%%%%%%%%%%

We have performed several numerical tests to illustrate problems
arising in multi-fluid flows (when simulated with {\PROMETHEUS}) and to
demonstrate the capability of the CMA method.

%%%%%%%%%%%%%%%%%%%%%%%
\subsection{Shock tube}
%%%%%%%%%%%%%%%%%%%%%%%

The first problem is the shock tube test problem originally proposed
by Sod \bracite{s78}. We have modified this problem to include three
passively advected fluids. The initial state for this problem (Fig.\
\ref{f:SODxi})
%%%%%%%%%%%%%%
\begin{figure}
%%%%%%%<%%%%%%%
%%%\picplace{5 cm}
%%% SODxi
%%%\includegraphics[bb =  71 438 481 675, height=5 cm, width=\columnwidth]{fig1}
\caption[]{ 
Initial distribution of fluid phases in Sod's shock tube problem: 
$X_1$ -- solid line; $X_2$ -- dashed line; $X_3$ -- dash-dotted
line. 
}
\label{f:SODxi}
%%%%%%%%%%%%
\end{figure}
%%%%%%%%%%%%
is,
\[
  \vec{U}(0{\le}x{\le}0.5,t=0)
= \pmatrix{ \rho \cr u   \cr p   }
= \pmatrix{ 1    \cr 0   \cr 1   },
\]
and
\[
  \vec{U}(0.5<x{\le}1.0,t=0)
= \pmatrix{ \rho  \cr u   \cr p   }
= \pmatrix{ 0.125 \cr 0   \cr 0.1 },
\]
with a discontinuous distribution of $X_1$,
\beqarrs
  X_{1}(x,t=0)
=
\cases
{
0.8, & $0{\le}x{\le}0.5$,\cr
0.3, & $0.5{<}x{\le}0.75$,\cr
0.1, & otherwise,
}
\eeqarrs
and oscillating mass fractions of the two other fluids,
\beqarrs
X_2 = 0.3 \sin^{2}(20 {\pi} x), \\
X_3 = 1 - X_2 - X_3.
\eeqarrs
The simulations have been performed with an ideal gas equation of
state with $\gamma = 1.4$, and an equidistant grid of 100 zones.
Reflecting boundary conditions were imposed on both ends ($x=0$ and
$x=1$) of the computational domain.

We obtained three sets of results for different code configurations:
original PPM code (Fig.\ \ref{f:SODxf}, top row),
%%%%%%%%%%%%%%
\begin{figure}
%%%%%%%%%%%%%%
%%%\picplace{10.2 cm}
%%% SODxf
%%%\includegraphics[bb =  56  63 581 675, height=10.2 cm, width=\columnwidth]{fig2}
\caption[] { 
Distributions of fluid phases (left) and deviations of the sum of mass
fractions from unity (right) in Sod's shock tube problem at time $t=1$
for three different runs: original PPM code (top), PPM with FMA
(middle), and PPM with sCMA (bottom).  $X_1$ -- solid line; $X_2$ --
dashed line; $X_3$ -- dash-dotted line.
}
\label{f:SODxf}
%%%%%%%%%%%%
\end{figure}
%%%%%%%%%%%%
PPM with FMA (middle section, $\varepsilon_\Sigma=10^{-3}$), and PPM
with CMA (bottom part) but with no special modifications of the
interpolation algorithm included (sCMA). The left panels of Fig.\
\ref{f:SODxf} show the distributions of the fluids ($X_1$ -- solid
line, $X_2$ -- dashed, $X_3$ -- dash-dotted). The right panels give
the deviations of the sum of the mass fractions from unity,
$\Delta_\Sigma$ (see Eq.\ \braref{e:delsig}), with the lower right
plot in Fig.\ \ref{f:SODxf} illustrating the truncation error of the
CMA method (Eqs.\ \braref{e:phi} and \braref{e:xscale}).

The comparison of mass fraction profiles of different models at $t=1$
(Fig.\ \ref{f:SODxf}) shows a good agreement for $x{\la}0.85$ except
for some small amount of clipping near the extrema of $X_2$ and $X_3$.
However, we find large differences near the jump in $X_1$ at $x
{\approx} 0.87$. The amplitude of the sinusoidal variation of $X_2$
and $X_3$ is significantly reduced in case of FMA, and the initial
discontinuities in $X_1$ are strongly smeared.  The original PPM
scheme and the sCMA scheme are much less diffusive. Both jumps in
$X_1$ are clearly separated and the profiles of $X_2$ and $X_3$ do
smoothly vary near $x {\approx} 0.87$.

The differences are even larger near the right boundary of the
computational domain (Fig.\ \ref{f:SODxf}). PPM strongly violates
condition \braref{e:sumx}. $\sum X_i$ deviates from unity by up to 6\%
(where $X_1$ is discontinuous). The errors are much smaller for FMA the
deviations only being of the order of $10^{-4}$ which is close to the
chosen value of $\varepsilon_\Delta$.  However, as already noted
above, the smaller error is bought at the cost of a degraded
resolution.  The sCMA method gave the best result. It does not only
advect the fluids with high accuracy, but it is also able to keep
$\sum X_i=1$ at the level of machine accuracy. The only imperfectness
one notices is some overshoot in the distribution of $X_1$ just to the
left of the larger discontinuity signaling the first sign of a need
for additional modifications of the interpolation scheme of the mass
fractions (especially near composition interfaces).

To observe the long term behaviour of the three schemes we continued our
simulations up to $t=400$ (more than 75\,000 steps with a Courant number
of 0.8). Figure \ref{f:SODdxn}
%%%%%%%%%%%%%%
\begin{figure}
%%%%%%%%%%%%%%
%%%\picplace{10.8 cm}
%%% SODdxn
%%%\includegraphics[bb =  36  63 483 711, height=10.8 cm, width=\columnwidth]{fig3}
\caption[]
{
Long-term behaviour of mean and extreme deviations from condition 
\braref{e:sumx} in Sod's shock tube problem. Top: original PPM; 
middle: FMA; bottom: sCMA.
}
\label{f:SODdxn}
%%%%%%%%%%%%
\end{figure}
%%%%%%%%%%%%
shows the evolution of the maximum negative and positive deviations
from condition \braref{e:sumx} recorded for each time step together
with the mean absolute value of the deviation from unity averaged over
all zones. The results for the original PPM method (top panel in Fig.\
\ref{f:SODdxn}) indicate large variations (in excess of 20\%) which
would certainly destroy any solution sensitive to chemical
composition.  When using FMA we observe a rapid rise of the maximum
error which levels off after slightly exceeding $2\
\varepsilon_\Delta$. The results obtained with sCMA show a slow growth
of the maximum and minimum deviations, which seem to saturate at later
times.

%%%%%%%%%%%%%%%%%%%%%%%%%%%%%%%%%%%%%%%%
\subsection{Two interacting blast waves}
%%%%%%%%%%%%%%%%%%%%%%%%%%%%%%%%%%%%%%%%

This test problem was originally proposed by Woodward \bracite{w82}.
It gained much of its popularity later when being used by Woodward
\& Colella \bracite{wc84} in their study of various advection schemes
in case of flows with strong shocks. In this problem the initial state
consists of a low-pressure region located in the central part of the
grid,
\[
  \vec{U}(0.1{<}x{<}0.9,t=0)
= \pmatrix{ \rho \cr u \cr p    }
= \pmatrix{ 1    \cr 0 \cr 0.01 },
\]
which is bounded by two regions of (different) high pressure
\[
  \vec{U}(0{\le}x{\le}0.1,t=0)
= \pmatrix{ \rho \cr u \cr p    }
= \pmatrix{ 1    \cr 0 \cr 1000 },
\]
and
\[
  \vec{U}(0.9{\le}x{\le}1,t=0)
= \pmatrix{ \rho \cr u \cr p   }
= \pmatrix{ 1    \cr 0 \cr 100 }.
\]
For our test runs we used three passively advected fluids with mass
fractions that are initially varying smoothly across the entire grid
(Fig.\ \ref{f:BWxi}),
%%%%%%%%%%%%%%
\begin{figure}
%%%%%%%%%%%%%%
%%%\picplace{5 cm}
%%% BWxi
%%%\includegraphics[bb =  71 438 481 675, height=5 cm, width=\columnwidth]{fig4}
\caption[]
{
Initial distributions of fluid phases in the interacting blast waves test
problem: $X_1$ -- solid line; $X_2$ -- dashed line; 
$X_3$ -- dash-dotted line.
}
\label{f:BWxi}
%%%%%%%%%%%%
\end{figure}
%%%%%%%%%%%%
%
\beqarrs
X_1 = 0.5 x^2, \\
X_2 = 0.5 \sin^{2}(20 x), \\
X_3 = 1 - X_2 - X_3.
\eeqarrs
Again, we use an ideal gas equation of state with $\gamma = 1.4$. The
grid consists of 400 equidistant zones. Reflecting conditions are
imposed at both grid boundaries.

The results of our simulations at $t=0.038$ are shown in Fig.\
\ref{f:BWxf}.
%%%%%%%%%%%%%%
\begin{figure}
%%%%%%%%%%%%%%
%%%\picplace{10.2 cm}
%%% BWxf
%%%\includegraphics[bb =  56  63 581 675, height=10.2 cm, width=\columnwidth]{fig5}
\caption[]
{
Distributions of fluid phases (left) and deviations of the sum of mass
fractions from unity (right) in the interacting blast wave problem at
time $t=0.038$ for three different runs: original PPM code (top), PPM
with FMA (middle), and PPM with sCMA (bottom).  $X_1$ -- solid line;
$X_2$ -- dashed line; $X_3$ -- dash-dotted line.
}
\label{f:BWxf}
%%%%%%%%%%%%
\end{figure}
%%%%%%%%%%%%

Since this time the initial distributions of the fluids are smooth we
do not expect to see any discontinuities in the distributions of the
mass fractions at later times. However, a discontinuity is created in
the $X_2$ and $X_3$ distributions at $x \approx 0.6$ when using the
original PPM and the sCMA method (left column, top and bottom panel of
Fig.\ \ref{f:BWxf}, respectively). No such discontinuity is present in
the FMA data (middle panel). We identify the creation of such spurious
composition interfaces with the actual failure of the unmodified
discontinuity detection procedure of PPM (see Section
\ref{s:cma}). FMA once again proves to be more diffusive than the
other two schemes: high-amplitude variations of $X_2$ and $X_3$ as
seen in the PPM and sCMA data for $0.6 {\la} x {\la} 0.8$ have
markedly smaller amplitudes when calculated with FMA. Moreover, there
is no trace of an extremum in $X_3$ at $x {\approx} 0.75$. In other
parts of the grid all three methods produce very similar results.

As in the case of Sod's shock tube problem the condition
\braref{e:sumx} is most strongly violated when the original PPM method
is used (upper left panel in Fig.\ \ref{f:BWxf}). The maximum
deviation of 2\% occurs in that region where the FMA results are
mostly affected by the use of an additional flattening procedure. On
the other hand, FMA violates condition \braref{e:sumx} at the level of
$\varepsilon_\Delta$ with a single pronounced maximum at the spurious
composition interface created in the other two schemes. The sCMA method
produces the most accurate results both during the initial phases of
the evolution and in the long term evolution (lower left panel in
Fig.\ \ref{f:BWdxn}).
%%%%%%%%%%%%%%
\begin{figure}
%%%%%%%%%%%%%%
%%%\picplace{10.8 cm}
%%% BWdxn
%%%\includegraphics[bb =  36  63 481 711, height=10.8 cm, width=\columnwidth]{fig6}
\caption[]
{
Long-term behaviour of mean and extreme deviations from condition
\braref{e:sumx} in the colliding blast waves problem. Top: original 
PPM; middle: FMA; bottom: sCMA.
}
\label{f:BWdxn}
%%%%%%%%%%%%
\end{figure}
%%%%%%%%%%%%
The maximum deviations from \braref{e:sumx} exceed 10\% for the
original PPM method and fluctuate between 2 and 3 times
$\varepsilon_\Delta$ in case of FMA.

%%%%%%%%%%%%%%%%%%%%%%%%%%%%%%%%%%%%%%
\subsection{Shock-contact interaction}
%%%%%%%%%%%%%%%%%%%%%%%%%%%%%%%%%%%%%%

The initial state for this problem is,
\[
  \vec{U}(0{\le}x{\le}0.1,t=0)
= \pmatrix{ \rho \cr u \cr p    }
= \pmatrix{ 1    \cr 0 \cr 1000 },
\]
\[
  \vec{U}(0.1{<}x{\le}0.5,t=0)
= \pmatrix{ \rho \cr u  \cr p    }
= \pmatrix{ 1    \cr -1 \cr 0.01 },
\]
and
\[
  \vec{U}(0.5{<}x{\le}1,t=0)
= \pmatrix{ \rho \cr u  \cr p    }
= \pmatrix{ 10^4 \cr -1 \cr 0.01 },
\]
with $X_1 = 0.2$ for $x<0.15$, $X_1 = 0.6$ for $x \ge 0.15$, and
\beqarrs
X_2 = 0.3\sin^{2}(30 x) + 0.001, \\
X_3 = 1 - X_1 - X_2.
\eeqarrs
The initial abundances with a composition interface between $X_{1}$ and
$X_{3}$ at $x=0.15$ are shown in Fig.\ \ref{f:SHCDSxi}.
%%%%%%%%%%%%%%
\begin{figure}
%%%%%%%%%%%%%%
%%%\picplace{5 cm}
%%% SHCDSxi
%%%\includegraphics[bb =  71 438 481 675, height=5 cm, width=\columnwidth]{fig7}
\caption[]
{
Initial distributions of fluid phases for the shock-contact interaction
problem: $X_1$ -- solid line; $X_2$ -- dashed line; 
$X_3$ -- dash-dotted line.
}
\label{f:SHCDSxi}
%%%%%%%%%%%%
\end{figure}
%%%%%%%%%%%%
%
Again an ideal gas equation of state with $\gamma = 1.4$ is used.  The
grid contains 400 equidistant zones. The left grid boundary is
reflecting, while a flow-in boundary condition is imposed at the right
grid boundary. The state of the inflowing gas is equal to that of the
gas located near that boundary at the initial time.

The initial conditions create a strong shock wave at $x=0.1$ which
propagates towards the left, hits the composition interface (initially
located at $x=0.15$) and then collides with the strong contact
discontinuity (initially located at $x=0.5$) that slowly moves to the
left. Upon interaction a pair of shocks is generated. 

Figure \ref{f:SHCDSxf}
%%%%%%%%%%%%%%
\begin{figure}
%%%%%%%%%%%%%%
%%%\picplace{10.2 cm}
%%% SHCDSxf
%%%\includegraphics[bb =  56  63 581 675, height=10.2 cm, width=\columnwidth]{fig8}
\caption[]
{
Distributions of fluid phases (left) and deviations of the sum of mass
fractions from unity (right) in the shock-discontinuity interaction
problem at time $t=0.045$ for three different runs: original PPM code
(top), PPM with FMA (middle), and PPM with sCMA (bottom).  $X_1$ --
solid line; $X_2$ -- dashed line; $X_3$ -- dash-dotted line.
}
\label{f:SHCDSxf}
%%%%%%%%%%%%
\end{figure}
%%%%%%%%%%%%
shows the distribution of the mass fractions together with the
deviations from the condition \braref{e:sumx} at $t=0.045$ after all
strong interactions have already taken place. All three methods give
comparable results in regions of pure advective flow ($x \ga 0.4$).
Towards the left follows a region with low-amplitude variations in the
distributions of $X_2$ and $X_3$, which is much more diffused when
calculated with FMA (middle left panel in Fig.\ \ref{f:SHCDSxf}). The
composition interface ($x \approx 0.25$) also seems to be smeared out in
case of FMA, but it remains sharp in the other two cases. Finally,
there is some overshoot in the distribution of $X_1$ which can be seen
just to the right of the composition interface in the sCMA data. The
interface is slightly broader when calculated with the sCMA method
than with the original PPM method primarily because of the smooth
rounded profile associated with the overshoot.

The analysis of deviations from condition \braref{e:sumx} (right
column in Fig.\ \ref{f:SHCDSxf}) confirms our conclusions from the
previous tests. Original PPM produces deviations of the order of a few
percent. Most of the extra diffusion used by FMA to keep deviations
small occurs in the region of interaction between the shock and the
contact discontinuity, and near the composition interface. It is in this
region where differences in abundances between FMA and the other two
codes are most apparent. The long term behaviour of the deviations
(Fig.\ \ref{f:SHCDSdxn})
%%%%%%%%%%%%%%
\begin{figure}
%%%%%%%%%%%%%%
%%%\picplace{10.8 cm}
%%% SHCDSdxn
%%%\includegraphics[bb =  36  63 483 710, height=10.8 cm, width=\columnwidth]{fig9}
\caption[]
{
Long-term behaviour of mean and extreme deviations from condition
\braref{e:sumx} in the shock-contact interaction problem.
Top: original PPM; middle: FMA; bottom: sCMA.
}
\label{f:SHCDSdxn}
%%%%%%%%%%%%
\end{figure}
%%%%%%%%%%%%
indicates that there is no tendency for deviations to become smaller
with time. Using the original version of PPM we find errors in the
several percent range (up to 10\%). The typical deviations for FMA are
between 2 and 3 times $\varepsilon_\Delta$.  There is some systematic
increase in deviations visible in case of sCMA at late times, but it
does not seem to be of any importance.

%%%%%%%%%%%%%%%%%%%%%%%%%%%%%%%%
\subsection{Supernova explosion}
%%%%%%%%%%%%%%%%%%%%%%%%%%%%%%%%

For our final ``numerical'' example we have chosen the shock
propagation during the post-bounce evolution of a core collapse
supernova. We consider the very early ($t{\le}3\,$s) stages of the
evolution, when the just born supernova shock begins to sweep through
the stellar layers just outside the iron core triggering nuclear
synthesis. We will focus our attention on the role which numerical
diffusion plays in the process of nuclear burning and on its impact on
the final chemical composition of the thermonuclear processed
material.

For these calculations we have used \PROMETHEUS\ in a version which
allowed us to consider those physical processes which play an
important role during the early phases of the shock propagation. The
gas is assumed to be a mixture of 14 nuclei (\nuc{1}{H}, \nuc{4}{He},
\nuc{12}{C}, \nuc{16}{O}, \nuc{20}{Ne}, \nuc{24}{Mg}, \nuc{28}{Si},
\nuc{32}{S}, \nuc{36}{Ar}, \nuc{40}{Ca}, \nuc{44}{Ti}, \nuc{48}{Cr},
\nuc{52}{Fe}, \nuc{56}{Ni}). An $\alpha$-network with 27 reactions
coupling 13 of these nuclei (all except \nuc{1}{H}) is used to
describe nuclear burning. The approximate equation of state includes
contributions from radiation, the 14 fully ionized Boltzmann gases,
and positron-electron pairs (included in an approximate way).
Self-gravity of the stellar envelope is taken into account as well as
the gravitational attraction of a central point source which mimics
the nascent neutron star ($M_{\mathrm{ns}} \approx 1.28\,$M$_{\sun}$).

The simulations have been performed in one spatial dimension assuming
spherical symmetry. At the inner edge of the grid a reflecting
boundary condition is imposed, while free outflow is allowed through
the outer boundary.  The inner boundary is situated at $r_{in} =
1.376\,10^8\,$cm corresponding to a mass coordinate of
1.317\,M$_{\sun}$, which is well inside the Si shell.  The
computational domain, which extends out to a radius of
$6.4\,10^{9}\,$cm, is covered by 1600 zones (corresponding to a
resolution of 40\,km) in our standard resolution run. Additional
simulations with up to 12\,800 zones (corresponding to a resolution of
5\,km) have been performed to study the convergence behaviour of
different advection schemes.

The initial model is a 15\,M$_{\sun}$ pre-collapse model of a blue
supergiant (S.~Woosley, private communication), which closely
resembles a progenitor model of SN~1987A (Woosley, Pinto \& Ensman
\cite{wpe88}). The density, velocity and temperature profiles of the
initial model are shown in Fig.\ \ref{f:sn_uri}.
%%%%%%%%%%%%%%
\begin{figure}
%%%%%%%%%%%%%%
%%%\picplace{14.40 cm}
%%% sn_uri
%%%\includegraphics[bb =  25 16 463 722, height=14.40 cm, width=\columnwidth]{fig10}
\caption[] 
{ 
Density, velocity, temperature and composition profiles for the most
abundant species (from top to bottom) of the 15\,M$_{\sun}$ progenitor
model.  The high temperature region in the innermost
($r{\le}3\,10^{8}\,$cm) part of the Si shell results from the
deposition of $10^{51}\,$ergs of internal energy in order to initiate
the explosion.
}
\label{f:sn_uri}
%%%%%%%%%%%%
\end{figure}
%%%%%%%%%%%%

In order to launch the supernova shock wave, we created a thermal bomb
by adding $E_{\rm tb} = 10^{51}\,$ergs in form of internal energy to
the innermost ($r \le 3\,10^{8}\,$cm) parts of the Si shell.  The
resulting shock wave propagates rapidly outwards heating up the
stellar matter. Simultaneously a much weaker reverse shock propagates
inwards. The temperatures and densities behind the supernova shock are
sufficiently high to trigger thermonuclear burning and the production
of new elements .  The resulting release of nuclear energy slightly
enhances the explosion energy. Whenever the outward propagating shock
crosses one of the composition interfaces of the progenitor star (see
Fig.\ \ref{f:sn_uri}), a weak reflected shock is created. The
reflected shocks move inwards reheating the matter, which has just
been processed by the supernova shock (see velocity profile in Fig.\
\ref{f:sn_urf}).
%%%%%%%%%%%%%%
\begin{figure}
%%%%%%%%%%%%%%
%%%\picplace{11.05 cm}
%%% sn_urf
%%%\includegraphics[bb =  25 180 463 722, height=11.05 cm, width=\columnwidth]{fig11}
\caption[] 
{ 
Density (top), velocity(middle) and temperature (bottom) profiles of
the progenitor at $t=3\,$s obtained with the CMA method and a grid
resolution of ${\Delta}r=10\,$km.  The supernova shock has reached the
helium shell and hence has already passed several composition
interfaces. At every interface a weak reflected shock is created,
which can be recognized in the velocity profile. The large density
jump near $r \approx 10^{9.5}\,$cm separates shocked gas that
initially formed the thermal bomb from matter in the stellar envelope.
}
\label{f:sn_urf}
%%%%%%%%%%%%
\end{figure}
%%%%%%%%%%%%
A strong contact discontinuity at $r \approx 10^{9.5}\,$cm
corresponding to the mass coordinate $M_{r} \approx 1.415\,$M$_{\sun}$
separates the shocked envelope gas from matter initially belonging to
the thermal bomb.

Figure \ref{f:sn_nxme}
%%%%%%%%%%%%%%
\begin{figure}
%%%%%%%%%%%%%%
%%%\picplace{11 cm}
%%% sn_nxme
%%%\includegraphics[bb =  46 109 463 722, height=11 cm, width=\columnwidth]{fig12}
\caption[]
{
Chemical composition of the ejecta obtained with the FMA method as a
function of mass coordinate at $t=0$, 100, 300, and 500\,ms (from top
to bottom). The grid resolution is 40\,km.
}
\label{f:sn_nxme}
%%%%%%%%%%%%
\end{figure}
%%%%%%%%%%%%
shows the evolution of the chemical composition obtained with the FMA
method at a resolution of 40\,km in a narrow mass range ($\sim
0.1\,$M$_{\sun}$) close to the outer edge of the thermal bomb. Nuclear
burning is most intense very early on ($t<100\,$ms) when matter is
still dense and hot. At later times the chemical composition does not
change appreciately with one exception.  Production of \nuc{44}{Ti}
only begins around $t=100\,$ms (a barely visible bump between 1.355
and 1.36\,M$_{\sun}$) and lasts for ${\simeq}300\,$ms.

In many aspects the evolution is similar when calculated with the CMA
method at the same grid resolution (Fig.\ \ref{f:sn_cxme}).
%%%%%%%%%%%%%%
\begin{figure}
%%%%%%%%%%%%%%
%%%\picplace{11 cm}
%%% sn_cxme
%%%\includegraphics[bb =  46 109 463 722, height=11 cm, width=\columnwidth]{fig13}
\caption[]
{
Chemical composition of the ejecta obtained with the CMA method as a
function of mass coordinate at $t=0$, 100, 300, and 500\,ms (from top
to bottom). The grid resolution is 40\,km.
}
\label{f:sn_cxme}
%%%%%%%%%%%%
\end{figure}
%%%%%%%%%%%%
However, some important differences also exist. With the CMA method
mixing between \nuc{16}{O} and other nuclear species at $M_{r} \approx
1.415\,$M$_{\sun}$ is greatly reduced. \nuc{36}{Ar} and \nuc{40}{Ca}
are clearly separated from oxygen at $t=500\,$ms. There is an
indication of a \nuc{28}{Si} interface near $M_{r} \approx
1.41\,$M$_{\sun}$. The transition region extends only over two zones
(over about 5 zones in case of FMA) and is accompanied by a small
amount of overshooting towards larger radii. This region is difficult
to model due to the presence of the strong contact discontinuity
separating the stellar envelope from the thermal bomb. Without
additional flattening and monotonization of the mass fraction profiles
(as described in section \ref{s:cma}) the overshooting of \nuc{28}{Si}
is suspiciously large. We note that overshooting of the most abundant
species near composition interfaces might be a common problem for
hydrodynamic codes (see, for example, Fig.\ 4 of Woosley, Pinto \&
Ensman \cite{wpe88}), and certainly deserves further investigation.

Another difference between the FMA and CMA results is the mass of
\nuc{44}{Ti} produced in the simulations, which seems to be quite
sensitive to the amount of numerical diffusion. Since titanium is
synthesized via the reaction $\nuc{40}{Ca} (\alpha,\gamma)
\nuc{44}{Ti}$ and since enhanced diffusion results in a smoother
distribution of calcium, we computed several models where the
interpolation profile for calcium was constructed assuming four
different constant flattening coefficients $f_{\rm Ca}$ (models
$f_{0.25}$, $f_{0.50}$, $f_{0.75}$ and $f_{1.00}$, respectively).  An
additional extreme model (CMAZ) was calculated where all mass fraction
profiles are flattened completely thereby imposing a maximum amount of
numerical diffusion. The results (Fig.\ \ref{f:sn_xn_9_10})
%%%%%%%%%%%%%%
\begin{figure}
%%%%%%%%%%%%%%
%%%\picplace{9.2 cm}
%%% sn_xn_9_10
%%%\includegraphics[bb =  76 203 454 722, height=9.2 cm, width=\columnwidth]{fig14}
\caption[]
{
Total mass of \nuc{40}{Ca} (upper part) and \nuc{44}{Ti} (lower part)
as a function of time obtained with different advection schemes. Solid
lines: CMAZ (thin), FMA (medium), CMA (thick). CMA results with
additional flattening for \nuc{40}{Ca} are shown by dashed lines for
$f_{\rm Ca}$ equal to 0.25 (thin), 0.50 (thick), 0.75 (long thin), and
1.00 (long thick).  The scale on the right side gives the masses
normalized to the respective final mass obtained with CMA.
}
\label{f:sn_xn_9_10}
%%%%%%%%%%%%
\end{figure}
%%%%%%%%%%%%
show that the amount of \nuc{44}{Ti} is smallest when using CMA
($5.0\,10^{-4}\,$M$_{\sun}$), that it increases linearly with $f_{\rm
Ca}$ and that the result of FMA ($1.22\,10^{-3}\,$M$_{\sun}$) is
recovered when the calcium profile is kept totally flat throughout the
simulation (model $f_{1.00}$).  In case of maximum numerical diffusion
(model CMAZ) the amount of titanium ($1.69\,10^{-3}\,$M$_{\sun}$) is
even larger and exceeds that obtained with CMA by a factor of more
than three. Table~\ref{t:mti44}
%%%%%%%%%%%%%
\begin{table}
%%%%%%%%%%%%%
\caption[]
{
Total masses of \nuc{40}{Ca} and \nuc{44}{Ti} (in units of M$_{\sun}$)
at $t=3\,$s
}
\label{t:mti44}
\begin{flushleft}
\begin{tabular}{lll}
\hline
\noalign{\smallskip}
model		&	\nuc{40}{Ca}	&	\nuc{44}{Ti}	\\
\noalign{\smallskip}
\hline
\noalign{\smallskip}
CMA		&	0.00718		&	0.00050		\\
f$_{0.25}$	&	0.00699		&	0.00067		\\
f$_{0.50}$	&	0.00684		&	0.00085		\\
f$_{0.75}$	&	0.00670		&	0.00105		\\
f$_{1.00}$	&	0.00658		&	0.00121		\\
FMA		&	0.00662		&	0.00122		\\
CMAZ		&	0.00665		&	0.00169		\\
WPE$^{\rm a}$	&	0.006		&	0.0002		\\
\noalign{\smallskip}
\hline
\end{tabular}
\end{flushleft}
\begin{list}{}{}
\item[$^{\rm a}$] Model 15A of Woosley, Pinto \& Ensman \bracite{wpe88}.
\end{list}
%%%%%%%%%%%
\end{table}
%%%%%%%%%%%
summarizes these results; data taken from model 15A of Woosley, Pinto
\& Ensman \bracite{wpe88} (who used a different mechanism to initiate
the explosion!) are also shown for comparison.

Figure \ref{f:sn_xrf}
%%%%%%%%%%%%%%
\begin{figure}
%%%%%%%%%%%%%%
%%%\picplace{11 cm}
%%% sn_xrf
%%%\includegraphics[bb =  46 109 463 722, height=11 cm, width=\columnwidth]{fig15}
\caption[]
{
Composition structure of the progenitor at $t=3\,$s: CMAZ (top), 
FMA (middle), CMA (bottom).
}
\label{f:sn_xrf}
%%%%%%%%%%%%
\end{figure}
%%%%%%%%%%%%
shows the composition profiles in the ejecta at $t=3\,$s for our three
basic models: CMAZ (top), FMA (middle), and CMA (bottom).  In CMAZ all
abundances change smoothly and no particular feature can be
recognized.  In CMA there exist composition interfaces of \nuc{40}{Ca}
($\log r \approx 9.36$), \nuc{16}{O} and \nuc{32}{S} ($\log r \approx
9.5$), and several discontinuities (in \nuc{28}{Si}, \nuc{32}{S},
\nuc{36}{Ar}, \nuc{40}{Ca}) at $\log r \approx 9.5$. Out of these only
a relatively weak discontinuity in the distribution of \nuc{36}{Ar}
can be recognized in the FMA model.

Finally, we have studied the convergence properties with respect to
the production of heavy elements in our three schemes. The results for
\nuc{44}{Ti} (Fig.\ \ref{f:sn_conv_x10})
%%%%%%%%%%%%%%
\begin{figure}
%%%%%%%%%%%%%%
%%%\picplace{4.7 cm}
%%% sn_conv_x10
%%%\includegraphics[bb =  59 438 481 675, height=4.7 cm, width=\columnwidth]{fig16}
\caption[] 
{ 
Dependence of the production of \nuc{44}{Ti} on the grid resolution.
The total \nuc{44}{Ti} mass is shown as a function of resolution at
$t=3\,$s for models CMAZ (open squares), FMA (open circles), and CMA
(full circles), respectively.
}
\label{f:sn_conv_x10}
%%%%%%%%%%%%
\end{figure}
%%%%%%%%%%%%
indicate that with CMAZ and FMA the production of titanium decreases
as the resolution is improved. More interestingly, the CMA results
depend only weakly on the resolution. This behaviour can be understood
if we realize that there exists a composition interface in calcium
near which titanium is formed, and that (as demonstrated by our
numerical experiments; Fig.\ \ref{f:sn_xn_9_10}) the production of
titanium grows with the diffusivity of the advection scheme. Once the
composition interface is resolved and properly handled by the code,
the final mass of titanium becomes practically independent of the
spatial grid resolution.

%%%%%%%%%%%%%%%%%%%%%%%%%%%%%%%%%
\section{Summary and conclusions}			\label{s:conclusions}
%%%%%%%%%%%%%%%%%%%%%%%%%%%%%%%%%

We have derived a numerical approach which guarantees that the sum of
mass fractions equals unity in simulations of multi-fluid flows with
higher-order Godunov-type methods. Unlike other commonly used
numerical methods, the proposed scheme preserves the conservative
character of the underlying advection scheme. We would like to stress
that even if the advection step is formally written in conservation
form, this does not necessarily imply that the scheme is conservative
in case of multi-fluid flows. This fact is often overlooked.

Modifications of the interpolation step are needed in higher-order
Godunov-type methods to reduce the numerical diffusion near
composition interfaces. These modifications are described together
with procedures that ensure the monotonicity of the scheme.

The Consistent Multi-fluid Advection method (CMA) is proposed as a new
method to accurately describe multi-fluid flows and is implemented in
the PPM-based hydrodynamic code {\PROMETHEUS}.  Since the advection
part of PPM is well tested for a single fluid, we have not considered
simple test problems with known solutions, like e.g., the linear
advection of square profiles in the mass fractions.  Instead we
investigated the behaviour of the CMA method in case of 1D test
problems with both smooth and discontinuous composition profiles
involving flows with strong hydrodynamic discontinuities. As for these
problems no analytical solutions are known, the correctness of the
proposed method has been demonstrated by means of a convergence
study. Although other methods converge to the same solution too, the
CMA method is the only one, which simultaneously guarantees the mass
constraint $\sum X_i = 1$, i.e. the sum of the mass fractions is
always and everywhere equal to one.

In order to demonstrate the advantage of the CMA method, we have also
studied a problem of astrophysical relevance, shock-induced
thermonuclear burning in a supernova explosion.  It is shown that
numerical diffusion near composition interfaces can change the
composition of supernova ejecta by a factor of a few. The abundance of
titanium is most severely affected in our test calculations. The
consequences of these findings for explosive nucleosynthesis
calculations should be explored in more detail.

%%%%%%%%%%%%%%%%%%%%%%%%
\begin{acknowledgements}
%%%%%%%%%%%%%%%%%%%%%%%%

We thank Stan Woosley for providing us with his unpublished
presupernova models. We also like to thank Konstantinos Kifonidis, who
provided software (reaction network) and helped with many critical and
useful comments during the course of this work.  TP was partly
supported by grant 2.P03D.004.13 from the Polish Committee for
Scientific Research. The major part of the simulations was performed
on the RISC cluster of the MPA, and on the SGI Power Challenge at the
Interdisciplinary Centre for Computational Modelling in Warsaw.

%%%%%%%%%%%%%%%%%%%%%%
\end{acknowledgements}
%%%%%%%%%%%%%%%%%%%%%%

%%%%%%%%%%%%%%%%%%%%%%%%%

%%%%%%%%%%%%%%%%%%%%%
%
%
%
%%%%%%%%%%%%%%
\end{document}